\documentclass[conference]{IEEEtran}
\usepackage{cite}
\usepackage{amsmath,amssymb,amsfonts}
\usepackage{algorithmic}
\usepackage{graphicx}
\usepackage{textcomp}
\usepackage{xcolor}
\usepackage{float}
\def\BibTeX{{\rm B\kern-.05em{\sc i\kern-.025em b}\kern-.08em
    T\kern-.1667em\lower.7ex\hbox{E}\kern-.125emX}}

\begin{document}
\title{A Recurrent Neural Network Based Patch Recommender for Linux Kernel Bugs\\}

\author{\IEEEauthorblockN{Anusha Bableshwar\IEEEauthorrefmark{1} and
Arun Ravindran\IEEEauthorrefmark{4}}
\IEEEauthorblockA{Department of Electrical and Computer Engineering \\
UNC Charlotte, Charlotte, USA\\
Email: \IEEEauthorrefmark{1}abablesh@uncc.edu,
\IEEEauthorrefmark{2}arun.ravindran@uncc.edu,
}
\and
\IEEEauthorblockN{Manoj Iyer}
\IEEEauthorblockA{
\textit{Canonical USA Ltd.}\\
Austin, USA \\
Email: manoj.iyer@canonical.com}

}

\maketitle

\begin{abstract}
Software bugs in a production environment have an undesirable impact on quality of service, unplanned system downtime, and disruption in good customer experience, resulting in loss of revenue and reputation. Existing approaches to automated software bug repair focuses on known bug templates detected using static code analysis tools and test suites, and in automatic generation of patch code for these bugs. We describe the typical bug fixing process employed in the Linux kernel, and motivate the need for a new automated tool flow to fix bugs. We present an initial design of such an automated tool that uses Recurrent Neural Network (RNN) based Natural Language Processing to generate patch recommendations from user generated bug reports. At the $50^{th}$ percentile of the test bugs, the correct patch occurs within the top 11.5 patch recommendations output by the model. Further, we present a Linux kernel developer's assessment of the quality of patches recommended for new unresolved kernel bugs.

\end{abstract}


\section{INTRODUCTION}

Software bugs in a production environment have an undesirable impact on quality of service, unplanned system downtime, and disruption in good customer experience, resulting in loss of revenue and reputation. More ominously, Linux kernel bugs result in vulnerabilities that are exploited by attackers to launch a variety of cyber attacks against computer systems. In this paper, we explore a new research direction that directly uses the bug report filed by users to recommend patches. Bug reports, and patch commit descriptions are manually generated by humans and expressed in a natural language such as English. Our goal is to relate a new bug description to the most closely related patch description that solved a similar bug in the past.

To highlight the motivation behind this new approach to designing patch recommender systems, we briefly outline the bug resolution workflow in a widely used Linux distribution. Bugs get reported in Launchpad/Bugzilla, typically by day-to-day users, and customers with service contracts. Linux as a whole experiences a very high volume and velocity of bugs reported, with hundreds reported monthly. The first step is to triage the the bug in order to identify the developer to assign the bug to, prioritize the bug, and then create tracks for various releases. Since production systems run stable (but older) versions of the Linux kernel, the assigned developer would look in the latest development releases of the kernel tree to check if this problem has been fixed upstream. If no fix was reported, he/she would then perform Git bisects between the target kernel version and the latest upstream development tree to identify the commit that resulted in the bug. Although there exists tools to automate the Git bisect process, much of this is a manual, time consuming process that needs experienced developers with domain knowledge on the subsystem that the bug is reported. Additional strategies include searching in forums or mailing lists to check if anyone else reported a fix. Furthermore, if the bug reported in the target kernel also exists in the latest mainline kernel, the developer would have to create a new patch, and get it accepted upstream. The developer would then backport the patch to all existing supported releases, especially if the bug fix is critical or a regression. Due to the time intensive process of fixing kernel bugs, and constraints on kernel developer time, many bugs take months to get patched, with some languishing for years without any resolution. An automated system that can leverage insights gained from bugs fixed in the past to shorten the bug fixing time for new bugs would therefore have numerous benefits including increased productivity of kernel developers, time and cost savings, enhancement of system security, and improvement of customer experience. We note that the same bug resolution bottlenecks identified above exists for the hundreds of user-level Linux packages as well. 

The current approach to automated bug fixing relies on static analysis tools and test suites to detect the bugs. As an example, Getafix, an automated bug fixing tool that Facebook uses in production, relies on the static analysis tool Infer, and the automated testing system Sapienz to detect bugs. \cite{FB18}. As such these tools are targeted at developers as a part of new code development flow. In contrast, in the Linux kernel, the user reported bugs are typically text descriptions of symptoms, along with kernel \textit{dmesg} output. We argue that to improve the efficiency of the kernel bug fixing process, an automated tool flow is needed that can parse user generated bug descriptions, and mine existing patches to enable developers to efficiently fix bugs.

In this paper, we explore the design of a patch recommender system for the Linux kernel that directly uses the bug report filed by users to recommend patches (bug-patch pair). Bug reports, and patch commit descriptions are manually generated by humans and expressed in a natural language such as English. Our goal is to relate a new bug description to the most closely related patch description that solved a similar bug in the past including upstream kernel versions. We note that using a simple keyword search is often insufficient, since the semantic context of the descriptions need to be understood to relate bug reports to patch commits. We explore the use of Natural Language Processing (NLP) to mine bug/patch descriptions. Recent years have seen huge gains in NLP due to the unprecedented success of Recurrent Neural Networks (RNNs). In particular, a family of RNNs known as Long Short-Term Memory (LSTM) networks has been highly successful in NLP tasks such as sentiment analysis, language translation, semantic similarity matching, and text summarization \cite{young2018recent}.

In this paper, we first create a bug-patch pair training data set after suitable pre-processing. We pose the bug-patch matching as a semantic similarity NLP problem. Note that unlike traditional semantic similarity problems such as detecting duplicate text, the bugs and patch descriptions are fairly dissimilar. After generating a custom word embedding for the bug-patch dataset, we train a Siamese LSTM Recurrent Neural Network based on 70\% the original bug-patch data set using the Keras-Tensorflow framework. The data set is augmented with mismatched bug-patch pairs - that is, where the bug and the patch are not related. We then evaluate our approach with bugs from the test set, and determine the top-K matches for the bug from all existing patches.  At the $50^{th}$ percentile of the test bugs, the correct patch occurs within top 11.5 patch recommendations output by the model. Further we evaluate the relevance of the patch recommendations for new unresolved kernel bugs. We are currently working on improving our results.

\subsection{Key contributions}
The paper makes the following contributions -
\begin{itemize}
\item To the best of our knowledge this is the first reported attempt to use NLP on bug and patch text descriptions to build a patch recommender system for the Linux kernel.
\item We have assembled a bug-patch labeled data set for the Linux kernel for use by other researchers.
\item We have demonstrated the design of a Siamese LSTM based recommender system for predicting closest matching patch for an input kernel bug.
\end{itemize}

\section{BACKGROUND}
\label{sec:background}
In this section, we present a brief description of previously reported work related to this research.

\subsection{Related Work}
\label{sec:relatedwork}
 Monperrus \cite{Monperrus2018} has provided a comprehensive review of the literature for automatic bug patching (also known as bug repair) as of 2018.  In general these approaches use test suites and static analysis tools to detect bugs, and then generate patches for these bugs through either known solutions for different bug patterns, or by mining existing patches for possible solutions. Recent projects include Prophet \cite{long2016automatic}, GetaFix \cite{FB18}, Deep Repair \cite{White17}, and SimFix \cite{Jiang2018}. In work related to mining bug description text, in the Deep Triage project \cite{mani2019deeptriage}, a RNN based-model is used to process bug reports so as to assign it to the appropriate developer. However, unlike our work, no attempt is made to recommend bug patches.

\section{MODEL DESIGN}
\label{sec:recommendersystem}
We use an MaLSTM model \cite{mueller2016siamese} to generate a match score between bug and patch descriptions. MaLSTM has a shared LSTM network (Siamese LSTM) with two inputs - the bug description and the patch description. The output is a scalar that indicates the degree of match between the bug and the patch (a real number between 0 and 1). The words in the text are represented using a word embedding matrix (See Section \ref{sec:wordembeddings}). The MaLSTM uses the LSTM network to read in word-vectors that represent each input text and employs its final hidden state as a vector representation for each description. The vector representation of these descriptions are used to calculate the negative exponent of the Manhattan distance between them as described in Equation \ref{eq:madist}.

\begin{equation}
    y = exp ( - | h^{(bug)} - h^{(patch)}|)
\label{eq:madist}
\end{equation}

Here y is the similarity metric, and $h_{bug}$ and $h_{patch}$ are the vector representations of the bug and the patch generated by the LSTM layer.

\begin{figure}
\begin{centering}
\includegraphics[scale=0.4]{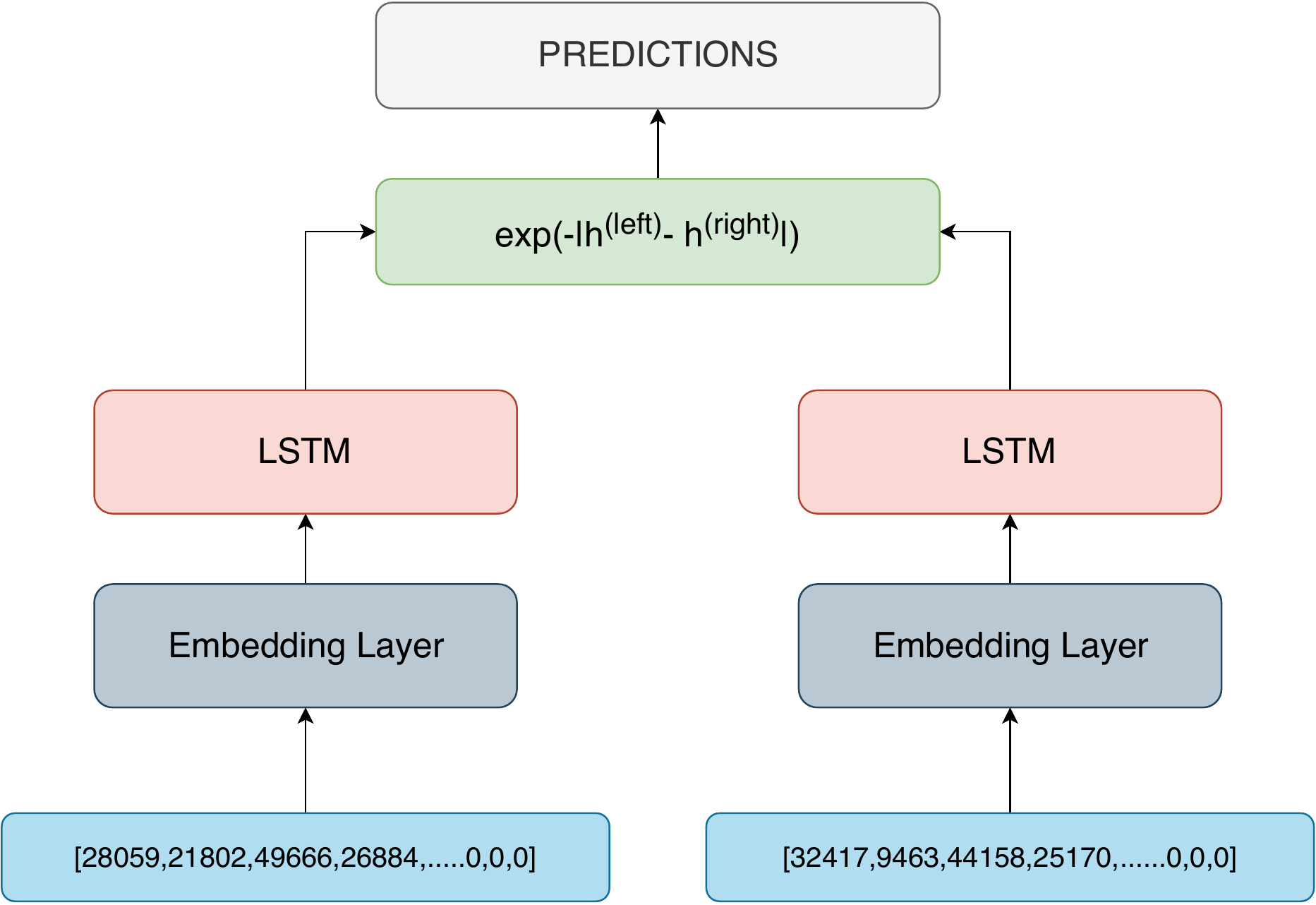}
\par\end{centering}

\caption[MaLSTM Architecture]{MaLSTM architecture. The text is represented using word embeddings in the embedding layer. Output from the shared LSTM layers (Siamese LSTM) is used to obtain a similarity representation between the bug and patch descriptions, which is then converted to a match score, a scalar value between 0 and 1 using a Manhattan distance based metric.}

\label{fig:malstm}
\end{figure}

Figure \ref{fig:malstm} shows the MaLSTM architecture. There are 4 layers in the model, the input layer, the embedding layer, the LSTM layer and the Manhattan layer.  The  duplicated layers on the left and right side carry the same network weights with the layers shared between the the two inputs.

The model is trained on the bug-patch data and drives the output (Manhattan distance) closer to the target by reducing the loss function after each training epoch. We minimize the negative log-likelihood of the observed labels at training time.

The match score output by the model is then used in the recommender system. A new bug is evaluated using the model against all existing patches, and the Manhattan distance based match scores are calculated. The patches are then sorted based on the match scores, and top-K patches are recommended as a possible fix to the bug. For practical usability reasons, K would be a small number between 3 and 5.

\section{DATA COLLECTION AND PREPARATION}
\label{sec:datacollection}
The bug-patch dataset is obtained from two open source platforms (Bugzilla \cite{bugzilla} and Linux Source Tree \cite{patches}). The data collected is a combination of bug-patch pairs with label 1, and bug-(non-patch) pairs with label 0. 

\subsection{Data scraping}
\label{sec:datascraping}
 For the analysis of the bug and patch, we scrape the short description part of the bug from Bugzilla, and patches from the Linux kernel git log. As a part of pre-processing the bug and patch descriptions, all special characters and symbols are removed. Then the text is tokenized and converted to lower case. Stopwords (most commonly occurring words such as a, an, the, they) are then removed from the text as they do not add much meaning to the descritpion. Python NLTK \cite{loper2002nltk} libraries are used to perform these tasks. 

\subsection{Word Embeddings}
\label{sec:wordembeddings}
Semantic vectors (also known as word embedding) allows the comparison of semantic meanings of the words numerically. Since there are Linux kernel specific jargon present in our data, a general pre-trained word embedding like \textit{GloVe} \cite{pennington2014glove} cannot be used.  We found that there were only 22\% of the words present in \textit{GloVe} from the bug and patch vocabulary. We use \textit{word2vec} \cite{mikolov2013distributed} to create our own word embedding to better represent the data. The corpus for training \textit{word2vec} is generated by selecting all the necessary words, followed by preprocessing the data and removing all stop words.

\textit{Word2vec} can be treated as a neural network with a single projection and hidden layer which we train on the corpus. The weights are used as the embeddings; the size of the hidden layer is equal to the dimension of the vector representation. The output of \textit{word2vec} is a vocabulary for each word having a vector along with it. The dimension of this vector is to be decided by the user; in our experiments we have used 100 as the vector dimension. 

\subsection{Data Preparation}
 We numerically index the words in the vocabulary starting from 0 to the length of vocabulary. The vocabulary is a list of all the unique words present in the entire text of dataset available. After creating a dictionary, with each word as the key and the number assigned as the value, the words in the text are replaced by these integer numbers. Any unknown words appearing here are given an $<UNK>$ token. The integer acts as a key for the words in the embedding matrix, and helps retrieve their vector representations. Both of these (the integer form of words, and the embedding matrix) are input to the embedding layer. The embedding layer replaces the integers with their vectors (n-dimensional) while training, and passes these to the LSTM layer.

\section{EVALUATION AND RESULTS}
\label{sec:evalresults}
In this section we describe the evaluation methodology of the proposed patch recommender system, and present results.

\subsection{Experimental Setup}
The bug and patch data is collected as described in Section \ref{sec:datacollection}.  In order to reduce generalization error, the data is augmented through additional negative samples. Our complete dataset consists of 27,670 samples with a bug having one matching patch and 5 non-matching patches. The dataset is divided into training (19,369), validation (4,151), and test (4,150). The input data is limited to 100 words per description; those which are shorter are padded with zeroes.

The implementation of our proposed approach is done using the Keras neural network library based on TensorFlow library. The training was done on Nvidia GeForce GTX 1060 GPU with 6 GB memory. The training for each epoch took about 1 minute, and the entire training process of 94 epochs with early stopping took about 1.6 hours.

The baseline of the MaLSTM model is two parallelly running LSTM layers each with 50 neurons, followed by the Manhattan layer which calculates the Manhattan distance based metric between the two vectors. The model is trained using the Adadelta optimizer.

The model testing is done by taking a test bug and evaluating the Manhattan distance based match score against all patches in our database (2132 patches) using the MaLSTM model. The patches are then sorted according to the match score, and the index of the matching patch is determined. Our goal is to ensure the presence of the matching patch in the top-K recommended patches, with a low K value. 

\subsection{Model evaluation}

\begin{figure}
\begin{centering}
\includegraphics[width=0.5\textwidth]{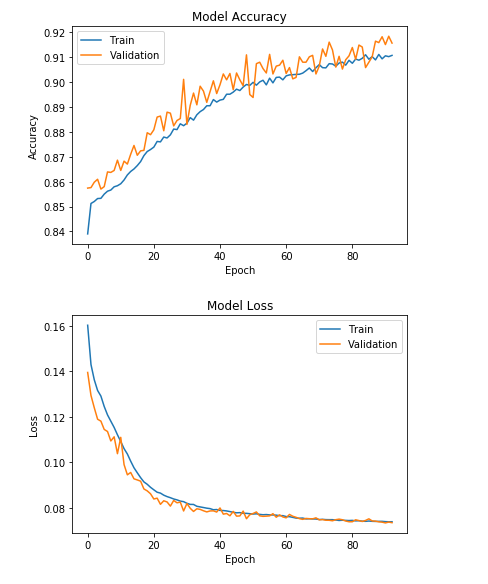}
\par\end{centering}
\caption[Accuracy and Loss during the training]{Accuracy and Loss of the training and validation data set.}
\label{fig:accuracyplot}
\end{figure}

We experimented with various numbers and sizes of the LSTM layer and found that 1 LSTM layer with 50 hidden units gave the best results thus far. As seen in Figure \ref{fig:accuracyplot}, the train and test accuracy as well as the loss show a monotonic decrease throughout the training process, indicating that the model is not overfitting over the training data. The model accuracy with the test set after training of the model was 90.8\%, after 94 epochs. This includes both true positives and true negatives. However, the goal of this model is not to predict whether the bug and the patch supplied to the model is a match or not. Instead the model serves as a means to calculate a match score used to recommend top-K patches for a new bug. To calculate the performance of the model, we chose a test bug and calculated the Manhattan distance based metric of that bug against all the patches existing in our database. Recommended patches with match score closer to 1 is hypothesized to be better fixes to the bug than the patches with match score closer to 0. 

\begin{figure}
\begin{centering}
\includegraphics[width=0.5\textwidth]{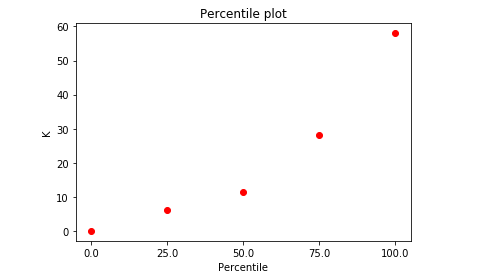}
\par\end{centering}
\caption{Percentile plot of the rank of the solution patch for test bugs}
\label{fig:percentileplot}
\end{figure}

To evaluate the performance of the patch recommender, 314 test bugs were evaluated, with each bug scored against all 2132 patches. Note that the test bugs were not used to train the model; however we know the solution patch for these test bugs. The goal was to determine the ranking of the solution patch among all 2132 patches.  The percentile plot for the ranking is shown in Figure \ref{fig:percentileplot}. The K value (rank) at the $50^{th}$ percentile (median) is 11.5, while at the $75^{th}$ percentile is 28.25. We note that the K-values we observe are too high to be practically useful. We are currently working on improving this number by enhancing the data set.

\subsection{Human evaluation}
To evaluate the quality of the recommended patches for new bugs for which no patches yet exist, 19 recently opened kernel bugs were obtained from Bugzilla. None of these bugs have been resolved yet at the time of writing of the paper. The top 10 patches to these bugs output by our recommender system, was evaluated by an engineer with many years of kernel bug fixing experience. The evaluations were done on a Likert scale of 1 to 5, where 5 (1) indicates strong agreement (disagreement) to the statement that one of the recommended patches is extremely useful in fixing the bug. We obtained scores ranging from 1 to 3 across 19 bugs. This is still a work in progress, and the scores are expected to improve with better model performance.

\section{CONCLUSIONS AND FUTURE WORK}
\label{sec:concl}

In this paper we have motivated the need to develop an automated bug fixing tool flow that can aid developer productivity by directly parsing user generated bug reports to recommend patches. We have proposed a Manhattan LSTM based network to generate a match score between the bug and the patch recommendations for the Linux kernel. While our model accuracy was high, the K value for the recommender system is still too large to be practically useful. We are currently working on improving our model by enhancing the training data set.

We consider the proposed approach that directly processes user bug reports, as a first step in the bug fixing process for complex software with a large user base. Future extensions of this work could seek to integrate code along with the text descriptions to yield ``ready to use" patches for the developer. Another line of research could focus on using encoder-decoder networks \cite{young2018recent} to ``translate" bug descriptions directly to patches.

\bibliographystyle{ieeetr}
\bibliography{references}

\end{document}